\documentclass[aps,prd,twocolumn,floatfix,nofootinbib]{revtex4}
\usepackage[dvipdfm]{graphicx} 
\usepackage{amsfonts}
\usepackage{amssymb}
\usepackage{amsmath}
\usepackage{bm}
\usepackage{graphicx}
\usepackage{multirow}
\usepackage{color}
\usepackage{graphics}
\usepackage{caption}
\usepackage{epsfig}
\usepackage{dcolumn}
\usepackage{rotating}
\usepackage{color}
\usepackage{hyperref}

\newcommand{\beq}{\begin{equation}}
\newcommand{\eeq}{\end{equation}}
\def\be#1\ee{\begin{align}#1\end{align}}
\newcommand{\ov } {\over }

\begin{document}

\title{Super-Planckian excursions of the inflaton and quantum corrections}

\author{Diego Chialva $^{1}$ and Anupam Mazumdar $^{2}$} 
\affiliation{
\vspace{1mm}
$^{1}$~Universit\'e de Mons, Service de Mecanique et gravitation, Place
du parc 20, 7000 Mons, Belgium\\ 
$^{2}$~Consortium for Fundamental Physics, Lancaster University,
Lancaster, LA1 4YB, United Kingdom} 


\begin{abstract}
Models of inflation with super-Planckian excursion seem
well in agreement with the recent observations of B- mode polarization
in the cosmic microwave background (CMB) radiation by the BICEP2 data.
In this note, we highlight the challenges faced by such models
from ultraviolet (UV) completion. 
In particular, we will discus radiative  corrections to the inflaton Lagrangian and to the gravitational sector.
We will emphasize why we would require an UV
complete theory  of gravity to tackle some of the issues for the super-Planckian excursion.
In particular, we will highlight how higher derivative terms in the inflaton and gravity sectors cause 
problems from non-locality and {\it ghosts}, if considered order by order, and thus prompt us to take into account 
infinite series of such terms. We will also stress how the presence of a scale of new
physics below the Planck scale would make some of the UV related problems more compelling 
and invalidate some of the remedies that have been proposed in the literature.
Finally, we will briefly speculate on possible ways of curing some of the challenges.

\end{abstract}


\maketitle

\section{Introduction}

The discovery of B-mode in the polarization of the 
CMB (cosmic microwave background) radiation 
at large angular scales by the BICEP2 team~\cite{Ade:2014xna} has made a very strong case for the
inflationary paradigm~\cite{inflation,Linde,Albrecht}. 
The signal is very well explained in terms of the primordial gravitational  
waves being stretched during inflation. The same stretching of the
modes, also acting on scalar (matter)
fluctuations, lead to the CMB temperature anisotropy, as observed by WMAP~\cite{WMAP} and
Planck~\cite{Planck-1}.

The amplitude of these tensor fluctuations is 
usually expressed by the
ratio of the scalar and tensor power spectra, 
${\cal P}_\zeta$ and ${\cal P}_T$, dubbed tensor-to-scalar
ratio, 
\begin{equation}
0.15\leq  r(k_\ast) \equiv \frac{{\cal P}_{T}(k_\ast)}{{\cal P_{\zeta}}(k_\ast)} \leq  0.27,
\end{equation}
at the pivot scale,  
$k_{\ast} = 0.002~{\rm Mpc}^{-1}$~\cite{Ade:2014xna}. 
The amplitudes of the matter power spectrum
${\cal P}_\zeta\sim 2.1\times 10^{-9}$ denotes the amplitude 
of the CMB temperature anisotropy~\footnote{Neglecting the subleading
  part due to tensor fluctuations.}. For a slow-roll dominated inflation, 
one can extract the potential energy of the inflationary vacuum: 
\begin{equation}
{\cal P}_T=\frac{2H_{inf}^2}{\pi^2 M_p^2}\approx \frac{2V_{inf}}{3\pi M_p^4}\sim 4.2\times10^{-10}\,.
\end{equation}
where $H_{inf}^2\approx (V_{inf}/3M_p^2)$ is the Hubble expansion rate
during inflation, $V_{inf}$ is the inflationary potential and  
$M_p\sim 2.44\times 10^{18}$~GeV is the reduced Planck mass. This
suggests that the inflationary
potential perhaps comes from the physics very  close to the Grand
Unified Theory scale, ie.  
$V_{inf}^{1/4}\sim 2\times 10^{16}$~GeV. This is also  the first
evidence of physical scale beyond the Standard Model and the scale of  
gravity at $M_p$. 

In fact, the simplest potential, also known as {\it Chaotic
inflationary} potential, for a single 
scalar field ~\cite{Linde,Linde-book}
 \begin{equation}\label{pot}
  V \sim V_0+ \frac{m_\phi^2}{2}\phi^2 +\cdots\,
 \end{equation} 
matches the current observations of CMB data extremely well, i.e. the amplitude of the
CMB temperature anisotropies power spectrum, its tilt, and
the tensor-to-scalar ratio at $r\sim 0.16$~\cite{Ade:2014xna}. 
The field $\phi$ could be either fundamental or
composite~\cite{Linde}.
The potential in Eq.~(\ref{pot}) yields an exponential inflation
within slow roll regime for $m_\phi \sim 10^{13}$~GeV, with a Hubble rate 
$H_{inf}\sim 10^{14}$~GeV.
In Eq.~(\ref{pot}), the ellipses stand for higher
order terms that would be present in the potential because of quantum
corrections, and 
are assumed to be
negligible in order for the model to work.  

Assuming such monotonic behaviour of the potential, one can derive an
upper bound on the value of $r$, known as the Lyth bound~\cite{Lyth}: 
 \begin{equation}\label{Lyth}
  r=16\epsilon\leq 0.003\left(\frac{50}{N}\right)^2\left(\frac{\Delta \phi}{M_p}\right)^2\,.
 \end{equation}
where $\Delta \phi$ denotes the field excursion during inflation,
lasted $N$ e-folds. For the observed $r\sim 0.2$, one
obtains $\Delta \phi\sim 10~M_p$. Here 
$\epsilon = (M_p^2/2)(V'/V)^2$ is the slow 
roll parameter, and the prime denotes the derivative of the potential w.r.t. $\phi$.

Both the slow-roll condition and the Lyth bound impose super-Planckian
field excursions for $\phi$ during inflation.
The challenge is to understand the {\it physical} consequences of this
super-Planckian VEV of inflation, see~\cite{Mazumdar:2010sa}. 

Indeed this is a {\it well-known}
problem~\cite{Linde-book,Riotto}, 
although the validity of the effective picture of inflation for the
potential in Eq.~(\ref{pot}) is justified, since
 \begin{equation}\label{energy}
  V_{inf}^{1/4} \ll M_p\,,
 \end{equation}
during the slow roll inflation,
the problem arises for the higher order corrections,
due to the fact that the field value of the
inflaton, i.e. $\phi \sim 10 M_p$, exceeds the Planck scale during inflation. 
This suggests that the effective field theory  treatment itself breaks down
at  the verge of super-Planckian inflation.  

In this note, we argue that
the quantum corrections will eventually require considering
higher-order terms in the Lagrangian, and in particular certain classes of
such corrections would spoil the  
inflationary potential obtained when the energy density is dominated
simply by the quadratic term in Eq.~(\ref{pot}). 

The article is organized as follows: we discuss the corrections to the
Lagrangian of the inflaton field, their magnitude, and the
relevance of having {\it different scales of high-energy physics} in section
\ref{LagCorSect}. We  
address some relevant arguments proposed in the literature to deal with these issues
in section \ref{IssRemSubs}. We then
focus on typically neglected corrections, affecting the kinetic terms for the inflaton and 
the purely gravitational sector, respectively, in sections  \ref{MattKineSect} and
\ref{GravCorrSect}. Finally,  in section \ref{RemeSect} we speculate on some
possible remedies to the problems that we have pointed out. 

\section{Higher order quantum corrections }\label{LagCorSect}

\subsection{Presence of multi-scales}

The higher order terms in the inflaton potential in Eq.~(\ref{pot})
would derive from the
coupling of the inflaton to heavy fields in the underlying UV-complete
theory~\footnote{In this work, we will focus on the quantum radiative corrections in
the inflaton Lagrangian, while neglecting {\it any} possible thermal corrections.}. 
Their presence follows naturally
from the facts, that

\begin{itemize}

\item{the existing inflation models are not UV complete.}

\item{the inflaton, being a gauge singlet 
order parameter, would couple to 
many degrees of freedom
in the complete theory, both hidden and 
Standard Model (SM) ones, see~\cite{Mazumdar:2010sa}.}

\item{gravitational couplings are universal, therefore  
one has to construct
an effective field theory argument based on the inflaton and its
interactions with other degrees of freedom, while taking into account 
gravitational corrections.}

\end{itemize}

One important point in our discussion is the role that different
energy scales would play. In fact, the question of effective theories in
relation with super-Planckian field excursions should not be
dealt with only in terms of $M_{p}$. 

One particular example of this, is (super)string
theory~\cite{Polchinski}, which we will often cite
as it is a well-developed candidate for a UV complete theory. Here, typically one would 
have a series of important scales: principally 
\textbf{\em a}) the compactification/Kaluza-Klein scale $M_c$, \textbf{\em b})
the string scale $M_s$, and \textbf{\em c})
the supersymmetry breaking scale $M_{\text{susy}}$.

The 
four-dimensional Planck scale
is a derived quantity related to $M_c$ and $M_s$, and
in the most well established models the scales are ordered as
$M_c \leq M_s \leq M_p$. 

\subsection{Radiative corrections due to inflaton couplings}

For the sake of simplicity and generality, let us consider a generic scale of new physics $M_f\leq M_p$, and we will discuss
cases, when $M_f\approx M_p$~\footnote{To be precise: corrections arising from graviton loops will
  always be weighted by $M_p$, while those coming from
  heavy fields will be suppressed by the scale $M_f$ relevant for
  those fields, where $M_f \leq M_p$.}. 
In this section, we will focus on the corrections arising from integrating out
fields coupled to the inflaton, in section \ref{GravCorrSect}
we will discuss how higher order correction do arise also
from the {\it purely} gravitational sector.

Generically for a SM gauge singlet inflaton, the inflaton can couple to 
various species of fermions and gauge fields:
 \begin{equation}
  V\sim \sum _{i}^{N}g_{i}\phi \bar\psi_{i}\psi_{i}\,,
  \qquad
  V\sim  \sum_{i}^{N} g^{\prime}_{i}\phi F^{i}_{\mu\nu}F^{i~\mu\nu}\,,
 \end{equation}
and, of course, gravity. Here, $g_{i},~g_{i}'$ are appropriate couplings. 

Such interactions of the inflaton to the matter field would lead to a host of
corrections both to the potential {\em and} to 
the kinetic terms for the inflaton: 
 \begin{equation}\label{gen-pot}
  \delta\mathcal{L} \sim \sum_{n}\lambda_{n} {\phi^{n} \ov M_{f}^{n-4}}
  + \sum_{n, m}d_{m}\Big({(\nabla\phi)^{2} \ov
    M_{f}^{4}}\Big)^m{\phi^{n} \ov M_{f}^{n-4}} + \ldots,
 \end{equation} 
where $(\nabla\phi)^{2} =g^{\mu\nu}\nabla_\mu\phi\nabla_\nu\phi$,
$\nabla$ being the covariant derivative for the metric $g_{\mu\nu}$,
and $\cdots$ indicates terms where more than one derivative act on the
inflaton field. The operators are
weighted by the scale $M_f$ at which the new physics appears, with
$\lambda_n,~d_n \sim {\cal O}(1)$ for $g_i,~g'_i \sim {\cal O}(1)$. 

The corrections to the kinetic terms have been somewhat
neglected in the literature regarding super-Planckian field excursion, see for
example~\cite{Riotto, Lyth:2014yya,Burgess:2014tja,
Baumann:2014nda}, disregarding the rather distinctive issues that they
would introduce as compared to the corrections to the
potential~\footnote{\label{HighDerMod} Models of higher-derivative inflations have been
  discussed, such as k-inflation, Galileon models, Horndeski-like
  theories~\cite{kinfla, Burrage:2010cu}, but in those cases the
  Lagrangians are truncated at a certain order, or have specific ad-hoc
  structures that eliminate the issues which we will point out in the
  following. These models are {\it not}
  UV complete, and also face the open problem of UV completion and
  robustness of the inflationary predictions against
  corrections when the Lyth bound applies, as for example for
  k-inflation~\cite{Baumann:2006cd}. We will also 
  discuss how the examples where the
  high-energy completion has been dealt with, namely DBI inflation
  \cite{DBIinf}, do not {\it address} the points we will raise, see section \ref{GravCorrSect}.}. 
In fact, they too play an essential
role as we will discuss in particular in sections \ref{MattKineSect} and \ref{GravCorrSect}.

In the case of string theory, the non-renormalizable operators 
in Eq.~(\ref{gen-pot}) come
from higher dimensional and string loops, and the corrections are suppressed by the scale
of heavy degrees of freedom~\footnote{We will not discuss light moduli and
  multifield scenarios in this paper.} and the string coupling $g_s$. 
In this setup, the problem of destabilizing the effective
four-dimensional picture for super-Planckian field
excursion becomes particularly relevant.

\subsection{Issues concerning radiative corrections and proposed remedies}\label{IssRemSubs}

In this section we will discuss the validity of some relevant
remedies proposed to cope with the issue of the higher
order corrections to the inflaton action in presence of
super-Planckian field excursion. 
In the following sections \ref{MattKineSect} and
\ref{GravCorrSect}, we will further focus on important unsolved issues that have
been typically overlooked in the literature.

\subsubsection{Small numbers}

One evident possibility for reconciling with the current
CMB observations is to require small couplings of the inflaton, 
i.e. $g,~g'\leq 10^{-3}$ or so,
so that the fine-tuning would lead to sufficiently small $\lambda_n,~d_n$ in Eq.~(\ref{gen-pot}).
For example, to maintain the flatness of the potential generating the right amplitude of 
scalar and tensor perturbations, one would require, $g^4\ll 10^{-12}$
for the $\phi^4$ term.

This is all right, and one can systematically make the higher order 
non-renormalizable couplings small order by order, although the fine
tuning might become very strong if the inflaton $\phi$
has super-Planckian excursion and $M_f < M_p$. However, this raises the
question why and how nature would generate these small numbers, 
leading to the anthropic arguments, which we will not resort to in this discussion~\footnote{There has
been a discussion of a somewhat technically natural suppression of the
couplings of light fields in some string theory setups~\cite{Burgess:2010sy}, 
 but  one still has to resort to the full UV-complete analysis, which is lacking~\cite{Burgess:2010sy, Burgess:2014tja}. Moreover,
in explicit calculations,  the corrections to kinetic terms do not
appear to be tunable \cite{Burgess:2010sy}. Also, other light fields could arise
besides the inflaton \cite{Burgess:2010sy}, which would spoil the cosmological predictions
as strict constraints arise from the {\it isocurvature
  perturbations}~ 
 \cite{Planck-1}. Finally, building up models in agreement with
  the BICEP2 results in those scenarios has been shown to be difficult
  (sometimes leading again to {\it anthropic arguments} to motivate the choice
  of compactifications and the mass spectra \cite{DiffLVSm, Burgess:2014tja}).}.

  \subsubsection{The case when $M_f = M_p$}

For the specific case, when $M_f = M_p$, it has been argued that
Eq.~(\ref{gen-pot}), obtained by integrating out heavy
fields using their standard propagators, would simply not be valid,
because of black hole formation.

Indeed, one might be able to argue that according
to the Einstein's theory, if the inflaton has a
super-Planckian VEV with $g\sim {\cal O}(1)$, the coupled
fermions and/or gauge fields would be forming a blackhole
of mass $m_{\psi, A_\mu} \sim g\langle \phi\rangle \sim 10M_{p}$ 
with Planckian-sized Schwarzschild's radius 
$r_s\sim (10 g / 8\pi M_p)$.  
Therefore, one might argue, as in Ref.~\cite{Dvali}, 
that the inflaton is {\it virtually} coupled to a sea of back holes,
and therefore the effective correction should have a  
Boltzmann suppression, and enter the potential as
 \begin{equation}
   V \supset {\cal N}~e^{-S}~{\mathcal{O}_i(\phi, \nabla\phi, \ldots) \ov M_{f}^{\Delta_i-4}},
     \qquad S={g^2\phi^2 \ov M_p^2}\,. 
 \end{equation}
for all possible operators $\mathcal{O}_i$, with dimensions $\Delta_i$, salvaging the perturbative expansion. 

However, a couple of points arise. The first one is that this argument
relies on the fact that there is {\em no} scale of new physics before $M_p$, the scale of
black hole formation~\cite{Dvali}. To assume the validity of simply the
Standard Model and General Relativity right up to $M_p$ is a rather strong
assumption. 

An interesting observation concerns the issue of back reaction when $M_f \approx M_p$. 
Just from the above arguments, the universe at the onset of
inflation would  be filled with black holes. The number of such
black holes  can be estimated by counting
the number of degrees of freedom that the inflaton interacts with. For example,
if the inflaton couples universally to $n_\ast$ fermion species, then the
total energy density stored in the sea of black holes for $O(1)$
couplings would 
be roughly of the order of $\sim (n_\ast m_\psi)^4\sim ( 10n_\ast g M_p)^4$, 
for $m_\psi \sim g\langle \phi\rangle$ and $\langle \phi\rangle \sim
10M_p$. This energy density should be less than the inflaton
energy density, which requires, taking into account BICEP2 results, 
$(10n_\ast gM_p )^4 < (m_\phi^2\phi^2)\sim 10^{64}~({\rm GeV})^4$, or
\begin{equation}
n_\ast \times g < 10^{-3}.
\end{equation}
In fact, the constraint from this counting argument
leads to a value of the coupling similar to the fine tuning required
to make the quantum corrections sufficiently small. Indeed, 
the $g\phi\bar\psi\psi$ interactions would yield a
$\lambda\phi^4$ term, which would be negligible for
$\lambda < 10^{-12}$, in turn requiring $g < 10^{-3}$.

Another related question is whether a
semiclassical treatment of gravity would be valid or not at $M_p$
during inflation - the required condition is that the total
energy density is sub-Planckian. Usually, Eq.~(\ref{energy}) is
claimed to guarantee this. 
However, Eq.~(\ref{energy}) does not take into account the full series of
corrections to potential and kinetic terms, see
Eq.~(\ref{gen-pot}). To ensure that the total energy density is
sub-Planckian in presence of super-Planckian field excursions
would require again fine tuning.

One can also decide to abandon the semi-classical approach to gravity,
but at that point one would require 
a full non-perturbative formulation of gravity, which we sorely lack
at this moment. 
Even in string theory, we do not have a full understanding of all
orders $\alpha'$ and loop corrections (we will come back to this point
in the following).

\subsubsection{Resumming the potential}

The non-renormalizable operators in the series in Eq.~(\ref{gen-pot}) 
generically come with different signs, hence the properties of
their resummation may not be evident from the individual terms.
For example, in the cases of purely gravitational corrections
(graviton loops) it has been shown that the corrections to the {\em potential}
can be resummed and the expression is weighted by 
the inflaton potential and its derivative w.r.t. the
inflaton field $\phi$, i.e. 
$V(\phi),~V^{\prime\prime}(\phi)$~\cite{Smolin:1979ca}.
These parameters are then considered under {\it slow roll conditions},
which demand that $V^{\prime}(\phi) \ll M_P^{3}, V(\phi) < M_P^{4}$, and
so on, see for instance~\cite{Linde-book}. 

However, there are a couple of points  which arise: 
\begin{itemize}
\item{ the very foundation of the argument is based on
the validity of the slow roll conditions, 
and the universe has begun inflating.
However, inflation is not guaranteed at the first point, 
if we consider corrections such as Eq.~(\ref{gen-pot}), contrary to what is being generally
argued, see sections \ref{MattKineSect}, \ref{GravCorrSect}}
\item{ it is not guaranteed that the sum of the other corrections to the potential, beside
the purely gravitational ones, lead to the same sort
of rearrangement in the final result and leaves intact the predictions
arising from the lowest-order terms. Even more remarkably, for what concerns
kinetic terms correction that is {\em not} the case, due to the appearance
of ghosts and singularities (see sections \ref{MattKineSect}, \ref{GravCorrSect}).}
\end{itemize}

There is one case where the corrections would
be expected to be naturally weighted by the tree-level potential,
giving rise to a suppression of higher terms during inflation: if the
tree-level potential softly breaks an underlying symmetry such as {\it shift
symmetry}, and the inflaton is a pseudo
Nambu-Goldstone boson. 

In that case higher corrections are also proportional to the
symmetry breaking terms, and, when these are small, the corrections can
be under control ~\cite{Kawasaki:2000yn,Kaloper:2008fb, Kaloper:2011jz, Kaloper:2014zba}. 
However, to avoid the breaking of continuous global symmetry operated by gravity,
these shift-symmetries should be related to underlying short-scale fundamental
symmetries such as gauge ones \cite{Dimopoulos:2005ac, Kaloper:2011jz},
and the corrections to the inflaton kinetic term are not guaranteed to
be negligible, see \cite{Kaloper:2011jz} and section \ref{RemeSect}.

\subsubsection{Stationary points in the potential}

One the other hand, one can always entertain a slightly different
approach: could the corrected potential lead to
inflation thanks to {\it stationary} or {\it inflection}-like points,
where the higher derivatives of the potential with respect to $\phi$ vanish?

Examples of inflection-point and saddle-point inflation where
respectively $V^{\prime\prime}(\phi_0)=0$ and 
$V^{\prime}(\phi_0)=V^{\prime\prime}(\phi_0)=0$ have been studied
in~\cite{Allahverdi:2006iq} and \cite{Allahverdi:2011su}.

For instance, one may consider a potential of the
type
 \begin{equation} \label{SimpleInflect}
  V= |f(\phi)|^2\,,~~~~f(\phi)\equiv \sum_{n>1}\lambda_{n}\frac{\phi^{3n-1}}{M_p^{3n-3}}\,,
 \end{equation}
where we assume the canonical kinetic term for $\phi$ for the time
being. 

The values
of  $\lambda_{n}$ could be kept arbitrary at the time being, but soon 
we will see that the cosmological observations would start putting
constraints on the lowest order coefficient. For $n_{\text{max}}=1$, the potential  
is renormalizable, and yields $\sim \lambda_1^2\phi^4$ term.  

Potential of type Eq.~(\ref{SimpleInflect}) can be analyzed in the
context of {\it catastrophe theory}. The highest order term is called 
the {\it catastrophe germ} and the coefficients of the lower orders are
called {\it control parameter}, dictating at which field values 
the critical and/or inflection points are located, see Refs~\cite{Allahverdi:2011su,Arnold}.  

For example, 
let us consider the case with $n_{\text{max}}=3$. Then
$$V^{\prime}(\phi) =f^{\prime}(\phi)f^{\ast}(\phi)+ h.c.$$ is zero at 
the points: $\phi= (0, a^{1/3}, b^{1/3})M_p$, where $a + b = -5\lambda_2/8\lambda_3$ and $ab = \lambda_1 /4\lambda_3$ . 
These solutions exist for any values of
$\lambda_{1},~\lambda_{2},~\lambda_{3}$ for $\phi$ being
complex~\cite{Allahverdi:2011su}.
In fact this argument goes beyond $n=3$, and beyond the specific case
of saddle points, i.e. those where $V'=V''=0$. In fact, as shown in
Ref.~\cite{Allahverdi:2011su}, one can find the roots where $n-1$ 
derivatives of the potential vanish. This is a
consequence of the fact that there are $n-1$ complex roots, $a_i$, of 
 \begin{equation}
  f^{\prime}(\phi)\propto \phi \Pi_{i=1}^{n-1}[(\phi/M_p)^3-a_{i}]\,.
 \end{equation}
When two roots coincide, one gets {\it inflection-points}, and for
higher degenerate points one requires more roots to coincide. 

However, a simple analysis of $n_{\text{max}}=3$ would illustrate that one still
needs the lowest order coupling to be very tiny. Indeed, 
the inflection point is given by:
 \begin{equation}
  \phi =\phi_0\exp(i\pi/3,~i\pi,~i5\pi/3)\,,~\phi_0=\left(\frac{5\lambda_2}{16\lambda_3}\right)^{1/3}M_p\,,
 \end{equation}
 In principle, one can accommodate some of 
the coefficients to have large values, such as $\lambda_2\sim 10^{12}$ and
$\lambda_3\sim 10^{9}$, or $\lambda_2\sim {\cal O}(1)$ and
$\lambda_3\sim 10^{-3}$, such that  
 $\phi_0\sim {\cal O}(10)M_p$. However, to explain the current CMB
perturbations the amplitude of the temperature 
 anisotropy, which is given by~\cite{Allahverdi:2006iq}: $\delta_H\sim V^{\prime\prime\prime}(\phi_0)N_{CMB}^2/30\pi H_{inf}$, 
 where $H_{inf}\approx (V(\phi_0)/3M_p^2)^{1/2}$ and $N_{CMB}\sim
 50-60$ e-foldings, would 
 require
 \begin{equation}
  \lambda_1\left(\frac{16}{5}\frac{\lambda_3}{\lambda_2}\right)^{1/3}\leq 10^{-8}\,.
 \end{equation}
 One can see that irrespective of $\lambda_2$ and $\lambda_3$, if one
 requires $\phi_0\sim {\cal O}(10M_p)$, one inevitably requires a
 small 
 coupling, i.e. $\lambda_1\leq 10^{-8}$~\cite{Allahverdi:2011su}. 

In conclusion, also driving inflation at
 special points does not help in fine tuning the couplings  
 or the self interactions of the inflaton. One still needs at least
 some of the couplings to be very small.
Note also that so far we have always assumed canonical kinetic term
for illustration, but this is a point that we will argue to be 
incorrect in section \ref{GravCorrSect}.

\section{The importance of higher-derivative terms}\label{MattKineSect}

In this and the next section, we will focus on the corrections to the
{\em kinetic terms} in the inflaton and in the gravitational
sector of the theory. In the literature for super-Planckian inflation
these terms have often been
neglected, by invoking a series of arguments. Instead, we are now going to discuss
their relevance.

A first argument to discard the terms involving derivative of the
inflaton, both spatial and temporal, is that chaotic inflation can only occur
if their contribution to the energy density is
smaller than the Planck scale. It is acknowledged that
naturally derivative terms {\it do} scale like the Planck mass (typically
it would be $\phi\sim {\cal O}(10)M_p$, 
$\dot\phi,\sim {\cal O}(M_p^2)$,
~$\ddot\phi \sim{\cal O}(M_p^3)\cdots$), but it
has been argued that:

\begin{itemize}

\item{it makes sense to speak of (semiclassical) inflation only in
inflating patches, where those terms have to be small (and the
probability for such patches do exist is variably estimated), see
~\cite{Linde-book,Linde:1993xx}.}

\item{often there is an attractor solution  that drives the inflaton to a
regime where its derivatives are small~\cite{Linde:1993xx,Mukhanov:2005sc} (in this
case the probability of having an inflationary patch is order 1).}

\end{itemize}

However, these two arguments are not guaranteed to hold at all when
higher order terms are included. The attractor mechanism, indeed, need
to be re-analyzed taking into account the effects of higher derivative
terms in the field equations. 
Let us remind that typically these terms 
will lead to {\it non-locality} and stability issues~\cite{Ostragradski}, and  
they can also introduce extra states, such as {\it ghost-like} ones
that violates unitary and/or make the vacuum unstable.
Even abandoning the attractor solution argu-
ment, and invoking only the random generation of an
inflationary patch, the probability of generation of such
patch needs to be estimated considering the quantum
corrected action for superPlanckian field excursions.

One particularly relevant implication of higher derivative terms is
that they would generally force to consider a complete series of such
terms~\footnote{For the case of Horndeski-like theories, see footnote
  \ref{HighDerMod}\,. We recall here that such theories are not UV complete, and
  so face the challenges of preserving their properties against radiative corrections.}.
In fact, a rather generic feature  of covariant ``finite-order''
higher derivative theories is the presence of new ghost states.
Let us consider a simple scalar field
theory model of the form:
 \begin{equation}\label{example}
  S=\int d^4x\ [\phi \Gamma(\Box)\phi -V_{int}(\phi)],
  \qquad \Box =g^{\mu\nu}\nabla_{\mu}\nabla_{\nu}
 \end{equation}
where $\Gamma$ is a finite polynomial function. In this case one
can always write $\Gamma$ as:
\be
\Gamma(-p^2) \propto (p^2+m_1^2)(p^2+m_2^2)\dots(p^2+m_n^2)\,.
\label{poly-prop}
\ee
In order for the theory to be non-tachyonic, all the $m_i^2$ have to
be positive and real, we are using the metric convention
$-+++$. Moreover, if there are at least two discrete 
single poles (say $m_1\neq m_2$), then at least one of them is ghost
like, i.e. one of the residues has to be negative~\cite{Biswas:2005qr}:
\begin{equation}
{1\over (p^2+m_1^2)(p^2+m_2^2)}\sim {1\over p^2+m_1^2}-{1\over p^2+m_2^2}
\end{equation}
A double pole can be represented as the convergence of two simple
poles with opposite residues, and suffers from similar
problems~\cite{Smilga,Ostragradski}.  Similar arguments follow for
higher 
order poles, rendering higher derivative theories of the form Eq.~(\ref{example}) inconsistent.

These considerations prompt to take into account a whole series of
perturbations, which could be ghost-free. Examples of the Lagrangian
they could originate have been studied in the context of string field
theory and related contexts, see~\cite{SFT,SFT1,BGS}. 
For instance, in the so-called p-adic string action~\cite{Biswas}, see
also \cite{SFT1},
one has
\begin{equation}\label{p-adic}
{\cal L}\sim \frac{M_s^4}{g_p^2}\left[-\frac{1}{2}\phi e^{-\frac{\Box}{m_p^2}}\phi  +\frac{\phi^{p+1}}{p+1}\right]\,,
\end{equation}
where $g_{p}^{-2}=g_s^{-2}(p^2/p-1)$ and $m_{p}^2=2M_s^2/\ln p$,
and one can evade some of the problems mentioned above, because the 
action is intrinsically non-perturbative in nature.
In this case the propagator is modified 
in such a way that it does not contain any poles~~\footnote{The propagator has a form 
of an {\it entire function}, which does not have any poles other than the essential singularities at the infinities.}.
In other words there are no physical ghosts states around the true vacuum. On the other hand, the kinetic term
does modify the UV behaviour of the theory, in particular
this example has been studied in the context of inflationary cosmology in  
Ref.~\cite{Biswas}. 

Certainly, inflationary slow roll conditions are now modified and one
has to find a full solution to the action. 
Modifications akin to Eq.~(\ref{p-adic}) have been proposed in the
gravitational sector, altering the UV properties of
gravity, see Refs.~\cite{Biswas:2005qr, BGKM}. We will now turn to this point and study the consequences.

\section{Higher-derivative corrections from gravity}\label{GravCorrSect}

A super-Planckian field excursion and the necessity to take into
account near-Planck-scale physics prompt to deal with
the corrections also in the purely gravitational sector of the theory.
Let us illustrate the case at the lowest order, up to 
${\cal O}(h_{\mu\nu}^2)$, where $h_{\mu\nu}$ is a small perturbation
around the  
Minkowski metric $g_{\mu\nu}=\eta_{\mu\nu}+h_{\mu\nu}$,
$\mu, \nu=0, 1, 2, 3$. We would expect higher derivative corrections
of the type:
%
 \begin{multline}\label{gen-grav} 
  {\mathcal L}_{\text{gr}} \sim  {R \ov 2}+R {\cal F}_{1}\left({\Box \ov M_f^2}\right)R+
   R_{\mu\nu}{\cal F}_{2}\left({\Box \ov M_f^2}\right)R^{\mu\nu}
 \\
 + R_{\mu\nu\lambda\sigma}{\cal F}_{3}\left({\Box \ov M_f^2}\right)R^{\mu\nu\lambda\sigma} + \ldots\ .
 \end{multline}
%
where,
 \begin{equation}
  {\cal F}_{i}(\Box/M_f^2)=\sum^{\infty}_{n\geq 0}f_{i,~n}\Box^{n}\,,~~\Box =g^{\mu\nu}\nabla_{\mu}\nabla_{\nu}.
 \end{equation}
In the case of string theory such corrections would descend from
higher-dimensional and string dynamics, which yield $g_s^2$
and $\alpha'= M_s^{-2}$ corrections.

Such terms, if considered order by order, would yield ghosts, as shown
in Ref.~\cite{Biswas:2005qr,BGKM}, similar to the example we   
have demonstrated in our simple scalar field theory. 

The higher
derivative corrections in the gravity sector would modify the 
{\it graviton propagator} by introducing extra propagating states, which would be ghost-like.
The {\it only} way one can tame the issue of ghosts is to make sure
that there must not be any extra pole other than that 
of the {\it massless graviton}. Therefore, any modification in the graviton 
propagator should be such that it not only recovers the normal GR in
the infrared, but should not violate unitarity at all. This would
eventually add constraints on ${\cal F}_{i}$: they cannot be
arbitrary, and their form should be such that the unitarity and covariance
are maintained all the way  
up to the $M_p$ scale. This problem has been studied extensively in
Ref.~\cite{BGKM,Biswas:2013kla}, where the form of ${\cal F}_{i}$ have been
constrained  
appropriately to make sure that the only propagating degree of freedom
remains the massless graviton around the flat space
time~\cite{BGKM}. The modified kinetic terms have then to be entire
functions (except for the massless graviton pole).

For a particular choice of ${\cal F}_{i}$, the gravitational sector behaves like being 
{\it asymptotically free}, i.e. the gravity becomes weakened at the UV~\cite{BGKM,Biswas:2013kla}: 
\begin{equation}
{\cal F}_3=0,~{\cal F}_{1}(\Box/M_f^2)=\frac{e^{-\Box/M_f^2}-1}{\Box}=-\frac{{\cal F}_2(\Box/M_f^2)}{2}.
\end{equation}

Of course these conditions change if we discuss physics
in different space time backgrounds, as in case of deSitter or
anti-deSitter backgrounds.
During inflation, we are indeed interested in the de Sitter case, and
the above constraints have to be revisited accordingly. 

However, the points are that:

\begin{itemize}

\item{ one would have to consider 
complete series of higher derivative corrections}

\item{ higher derivative corrections in
the gravitational sector induces corrections into the scalar degree of
freedom. For instance,  
let us first concentrate on a sub-class of the above action, which was first considered in Ref.~\cite{Biswas:2005qr}:
\begin{equation}\label{sub-class}
 {\cal L}\sim R + R\sum_{i}c_{i}\Box^{i} R\,.
\end{equation}
This action is actually equivalent to a scalar-tensor action of the form~\cite{Biswas:2005qr}:
\begin{equation}
{\cal L}\sim \Phi R+\psi\sum_{I=1}^{\infty}c_{i}\Box^{i}\psi-{\psi (\Phi-1)-c_0\psi^2}\,,
\end{equation}
the equivalence can be seen by $\psi=R$ and $\Phi=1$.}

\end{itemize}

Incidentally, this action, Eq.~(\ref{sub-class}), has been studied extensively in the context
of cosmology. It yields a non-singular bouncing cosmology with the scale factor:
\begin{equation}
a(t)= \cosh\lambda t\,,
\end{equation}
where $\lambda \sim \sqrt{\rho_\phi/3M_p^2}$, with $\rho_\phi \sim
{\rm const}$, which signifies an era of super-inflation as pointed  
out in Ref.~\cite{BA}. The important lesson for us is to keep in mind
that the higher derivative terms in the gravitational sector tend to
ameliorate the UV properties  
of Einstein's general relativity.

The necessity to take into considerations also the gravitational
corrections is also important in the context of string theory. In fact,
string theory can resum all $\alpha' = M_s^{-2}$ corrections in
certain cases, leading for example to the DBI action \cite{Fradkin:1985qd} (which has been
used in the inflationary models of \cite{DBIinf}). 
However, this action does not include string loops, which are those
accounting for the higher-order gravitational corrections. At this
moment there is no UV complete formulation of string theory capable of
taking into account all $g_s$ and $M_s$ corrections, and the tools
coming from on-perturbative approaches such as gauge/gravity duality
are not sufficiently developed either.

\section{Possible remedies}\label{RemeSect}

Given some of the challenges we have posed for super-Planckian inflation, 
it is important to suggest some solutions to the problem, which would
be able at the same time to explain the cosmological observations such as the temperature 
anisotropy in the CMB and the BICEP2 data.

\begin{enumerate}

\item {\bf Sub-Planckian field excursion }\\
{\it \underline {Assisted inflation}:} It has been known for some time that inflation could be driven collectively 
by $N$ independent copies of the inflaton field, known as assisted inflation~\cite{Liddle:1998jc}, and its 
generalisation when specific cross-couplings were introduced~\cite{Copeland:1999cs}.
The simplest choice will be to take then $N$ copies of $m^2\phi^2$
potential. Similar constructions utilizing axion field have been
studied in ``N-flation'' \cite{Dimopoulos:2005ac}.

One of the virtues of such $N$ copies is that inflation can now occur at sub-Planckian VEVs as 
pointed out in Refs.~\cite{Kanti}, and in Ref.~\cite{Jokinen:2004bp}. However, in order to explain the 
temperature anisotropy one would require large number of fields: $N\sim
10^{3}-10^{4}$. This models can also account for 
$r\sim 0.16$, yielding predictions similar to a chaotic inflation model with super-Planckian VEVs.

In~\cite{Dimopoulos:2005ac}, this mechanism employing $N$ copies of
inflation has been studied in the string theory context with
axionic fields enjoying shift symmetry. One advantage of having $N$ fields
is that inflation can occur with the decay
constants $f_i$ of the axions 
below $M_p$, which solves
the problems of being able to engineer $f > M_p$, as would be needed in single field
models, see \cite{Banks:2003sx}. One drawback is that
the large number of fields can enhance the radiative corrections.

{\it \underline{Single field, inflection-point inflation}:} As an
alternative, one may imagine that $M_f=M_p$,  and realize inflation
via  
some {\it inflection-point} inflation~\cite{Allahverdi:2006iq,Allahverdi:2011su}.
We have briefly discussed such a model in the context of super-Planckian excursion, but 
the initial motivation for these models were to drive inflation within sub-Planckian field VEVs.
In principle it is possible to obtain large observable $r\sim 0.2$ with sub-Planckian excursion 
of the inflaton field if there exists an inflection point as pointed out in Refs.~\cite{SM}. One 
of the advantage is that both the energy density and the VEV of the inflaton remains below the 
cut-off for the validity of an effective field theory. 

However, in presence of a new fundamental scale $M_f< M_p$ and
sufficiently large field excursions, one has to revise the model 
taking into account higher derivative corrections in the inflaton and
gravity sector.

\item {\bf Asymptotically Free Gravity: non-singular bounce/loitering universe:}\\
There are hints that string theory motivated $\alpha'$ corrections in the gravitational 
sector might lead to ameliorating the ultraviolet aspects of gravity~\cite{Siegel:2003vt}, and Refs.~\cite{Biswas:2005qr,BGKM}.
The gravity can be weakened at scale close to the string scale, $M_s< M_p$, in such a way that it can resolve the singularity 
arising in the mini Schwarzschild's black hole (linearized solutions
of Eq.~(\ref{gen-grav}).  In such a case, even if the inflaton VEV 
exceeds that of $M_p$, one would not form Planckian size black holes~\cite{BGKM}, through inflaton coupling to the matter fields. 
Furthermore, such a non-perturbative formulation of gravity can also yield a non-singular bouncing 
cosmology~\cite{BGKM}. The sub-class of action shown in Eq.~(\ref{sub-class}) also yield a non-singular bouncing 
cosmology~\cite{Biswas:2005qr}, whose perturbations around the bounce
solution have been found to be stable~\cite{BMS}. 

One interesting point when weakening gravity is the possibility to
realize a prolong phase of {\it loitering universe} with  
a string gas domination~\cite{Brandenberger,Biswas:2006bs,Brandenberger:2014faa,Biswas:2014kva}.  
All these models do deviate from inflation to generate density
perturbations for the structure formation, and they still need a
mechanism to stretch the perturbations on super-Hubble scales.

\item {\bf Shift-symmetry from an underlying fundamental symmetry:}\\
As we have mentioned, a shift symmetry of the inflaton field would prevent the appearance
 of many higher-order terms. This approach has been employed in a
 number of inflationary models and reconsidered after BICEP2 results, see for example ~\cite{Adams:1992bn,Kawasaki:2000yn,
 Dimopoulos:2005ac, MonodromyInfl, Kallosh:2014vja,Hebecker:2014eua, Blumenhagen:2014gta}.
 In particular, if the
 shift symmetry descends from an underlying short-scale fundamental symmetry 
(such as it is the case in certain inflationary realisation where it
 derives from higher-dimensional gauge symmetries), it would not be
broken by gravitational effects, see~\cite{Dimopoulos:2005ac, Kaloper:2011jz}.
In the chaotic inflationary scenario the 
 shift symmetry is of course broken already by the $m_\phi^2 \phi^2$
 term, but this would be just a soft breaking
 term~\cite{Kawasaki:2000yn}, so that the higher 
 order terms would still be weighted by the soft breaking
 coefficients. 

 However, these models still present serious issues
 to provide predictions in agreement with the observations (for
 example, some of them need to accommodate for dimensionful
 parameters larger than $M_p$, or suffer from poor control of string
 loop corrections, or resort to anthropic arguments applied to the
 landscape of string flux compactifications to have certain hierarchies among
 scales or to tune certain coefficients, see the above references).
 Together with these issues, one should also investigate the corrections to the
 inflaton kinetic term. In fact, higher derivative kinetic terms do
 preserve the shift-symmetry, and would appear naturally from $g_s$
 and $\alpha'$ corrections.

 \end{enumerate}

In conclusion, super-Planckian field excursions of the inflaton would provide us with
an important handle on very high-energy physics. In particular, we
have pointed out that if the fundamental scale of nature is such that
$M_f \leq M_P$, one would be forced to consider  quantum  
corrections to the inflaton potential and to the kinetic term of the inflaton field.
Both these corrections are non-trivial, but in particular the higher derivative corrections
to the inflaton kinetic terms lead to ghosts and one has to sum these higher derivative operators 
to all orders to avoid the ghost problem.

Excluding an anthropically motivated fine tuning of the inflaton coupling to matter, 
we have outlined some promising approaches, which, interestingly, appear to invoke specific features of the UV
theory, in terms of symmetries or
UV aspects of gravity. This further stresses the relevance of
fully understanding the physical implication  of the picture of
super-Planckian field excursions in the context of 
high-energy physics.

\acknowledgements
AM is supported by the Lancaster-Manchester-Sheffield Consortium for Fundamental Physics under STFC grant ST/J000418/1.
The work of DC is supported by the Belgian National ``Fond de la
Recherche Scientifique'' F.R.S.-F.N.R.S. with a contract ``charg\'e de
recherche''. AM would like to thank Tirthabir Biswas and Masahide Yamaguchi for clarifying some of the relevant issues. 


\end{document}